\documentclass{article}
\usepackage{spconf,amsmath,graphicx}

\usepackage{enumitem}
\setlist{nosep, leftmargin=14pt}

\usepackage{mwe} 
\usepackage{url}

\title{A Semi-automatic Cranial Implant Design Tool Based on Rigid ICP Template Alignment and Voxel Space Reconstruction}
\name{Michael Lackner, Jan Egger, Jianning Li}
 \name{Michael Lackner$^{1,2}$ \qquad Behrus Puladi$^{3,4}$ \qquad Jens Kleesiek$^{1}$ \qquad Jan Egger $^{1,2}$ \qquad Jianning Li$^{1}$}
 \address{$^{1}$ Institute for Artificial Intelligence in Medicine, University Hospital Essen (AöR) \\
     $^{2}$ Institute of Computer Graphics and Vision, Graz University of Technology \\
     $^{3}$ Department of Oral and Maxillofacial Surgery, University Hospital RWTH Aachen \\
     $^{4}$ Institute of Medical Informatics, University Hospital RWTH Aachen}

\begin{document}
%
\maketitle
\begin{abstract}
In traumatic medical emergencies, the patients heavily depend on cranioplasty - the craft of neurocranial repair using cranial implants. Despite the improvements made in recent years, the design of a patient-specific implant (PSI) is among the most complex, expensive, and least automated tasks in cranioplasty. Further research in this area is needed. Therefore, we created a prototype application with a graphical user interface (UI) specifically tailored for semi-automatic implant generation, where the users only need to perform high-level actions. A general outline of the proposed implant generation process involves setting an area of interest, aligning the templates, and then creating the implant in voxel space. Furthermore, we show that the alignment can be improved significantly, by only considering clipped geometry in the vicinity of the defect border. The software prototype will be open-sourced at \url{https://github.com/3Descape/Cranial_Implant_Design}.
\end{abstract}
\section{Introduction}
\label{sec:intro}

\begin{figure*}[htb]
\begin{minipage}[b]{1\linewidth}
  \centering
  \centerline{\includegraphics[width=18cm]{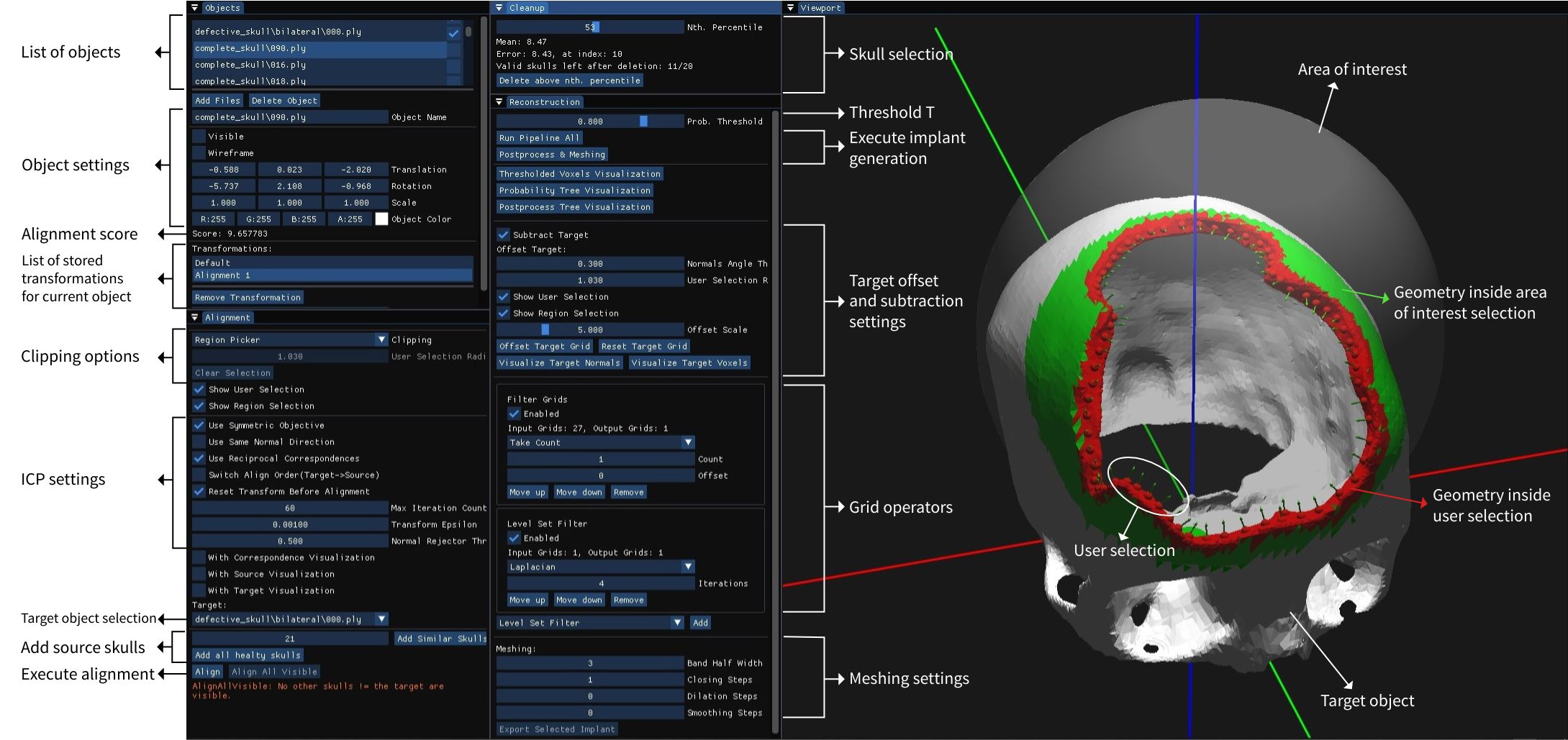}}
\end{minipage}
\caption{An overview of the user interface and the available settings of the software prototype. Most of the settings are optional.}
\label{fig:ui}
\end{figure*}

Reasons for the need of neurocranial surgery and repair, i.e., cranioplasty, include decompression surgery for stroke, craniectomy for intracerebral infections and epilepsy, as well as head trauma \cite{CranioplastyWithPCI, TherapeuticEffect}.
Primarily, cranioplasty aims to restore the functional and aesthetic aspects of the cranial (bone) structure, to protect the brain and improve the quality of life for the patiens \cite{CranioplastyWithPCI, TherapeuticEffect}. Compared to a synthetic allograft, the autogenous bone graft (autograft) is still considered the first choice for cranioplasty, due to  better reintegration and immune response, which however might not always be feasible due to damage, infection or unavailability \cite{CranioplastyWithPCI, IntervalWithPCI, RecentRevolution}. Yet, various studies \cite{CranioplastyWithPCI, IntervalWithPCI, RecentRevolution, PredictingReducingInfections, RiskFactorsOfAsepticBoneResorption} suggest the advantages of allocraft patient-specific implants (PSIs), over autografts, considering factors like accumulated cost, length of hospital stay, success rates, and alike \cite{IntervalWithPCI}. Furthermore, PSIs are a critical backup solution in cases where autogenous implants need to be removed due to infection or resorption, which are two of the most common complications, or if the size and shape of the defect do not allow for autografts \cite{TherapeuticEffect, IntervalWithPCI}. 

The application of modern manufacturing techniques such as computer-aided design, numerical control, additive and subtractive manufacturing in conjunction with computed tomography, massively streamlined the creation process of PSIs in recent years, making it a common choice for cranioplasty \cite{CranioplastyWithPCI, IntervalWithPCI, RecentRevolution}.
Nevertheless, the high production costs, availability, and the complexity of the process involved hinder, at least partially, the broad adoption of PSIs \cite{IntervalWithPCI, RecentRevolution, ComputerAidedPlanning}. However, the work of Lannon et al. \cite{CostEffectiveCranioplasty} shows that using cheap and broadly available technology, such as 3D printing, cuts down costs of production dramatically. Apart from that, researchers in the area of regenerative medicine and tissue engineering investigate how PSIs can be improved in aspects like immune reaction and integration, which might bring PSIs on pair with autografs in the future \cite{RecentRevolution}. Yet, further research is still needed, especially with regards to semi- and fully-automated implant creation tools, as current solutions usually require a high level of expertise, manual intervention, and time, which makes the design process one of the most complex, expensive, and least automated tasks in the whole process \cite{ComputerAidedPlanning, AutomaticSkullDefect}. 

As an attempt to solve the problem, we created a software prototype specifically tailored for semi-automatic creation of PSIs. On a high level, the implemented implant generation process involves setting an area of interest, aligning the healthy skulls to the damaged target, and then reconstructing the defect area by means of averaging and thresholding in voxel space. However, instead of considering the entire mesh as part of the alignment, we propose to only consider clipped geometry in the vicinity of the defect's boundary to improve the alignment in critical areas. The software is open-sourced to facilitate future research in this direction.

\section{Methods}
\label{sec:methods}

\subsection{Iterative Closest Point}
The main purpose of the Iterative Closest Point (ICP) algorithm is to perform rigid registration between two objects in 3D-space \cite{icp_method_for_registration}. As the name suggests, ICP estimates the optimal pose of the moving object relative to the fixed object in an iterative manner. At each iteration of the algorithm, nearby correspondence pairs between the two objects are found. The ideal transformation of the moving object is then calculated such that the total distance between the two objects is minimized. These steps are repeated until the mean squared distance is smaller than a given threshold, a maximum number of iterations has been reached or some other stopping criteria is fulfilled \cite{icp_method_for_registration}. In practice, more complex objective functions, such as point-to-plane or symmetric, are used,  since they usually converge faster than the point-to-point objective \cite{icp_symmetric}.

\subsection{Application Prototype}
The application, whose main user interface is shown in Figure \ref{fig:ui}, is written in C++ and used an OpenGL backend to render the ImGui user interface and 3D viewport. Additionally, we take advantage of the OpenVDB library for voxel processing and the Point Cloud Library for the ICP and Globally Aligned Spatial Distribution (GASD) implementations. To ensure fast processing and rendering during use, we additionally cache the dataset as OpenVDB grids, GASD features, and meshes on the first startup.

\begin{figure}[tb]
\begin{minipage}[b]{1\linewidth}
  \centering
  \centerline{\includegraphics[width=8.0cm]{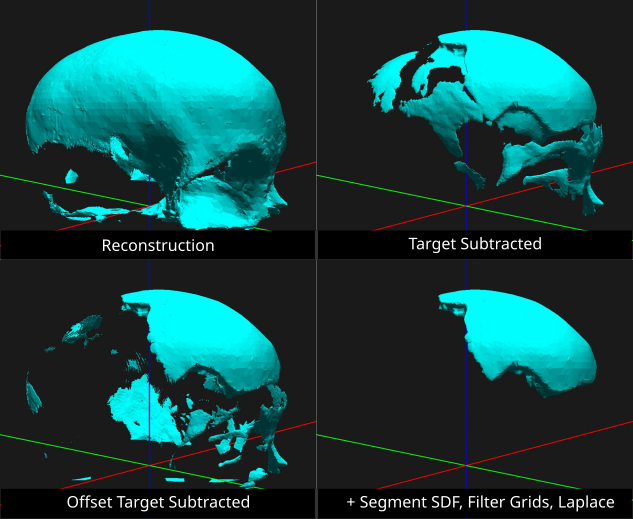}}
\end{minipage}

\caption{In the top right image we see that a lot of the artefacts are still connected with the implant after subtraction. When applying an offsetting to the target before the subtraction (bottom left) the artefacts are disconnected and can therefore easily be removed with the grid operators (bottom right).}
\label{fig:grid_operators}
\end{figure}

The overall workflow of the prototype application is described as follows. First, the user specifies an area of interest by placing a sphere over the defect area. In the following ICP alignment step we then use this information to clip the geometry outside of the selection. This way, we achieve a better alignment of the templates in the vicinity of the defect and therefore a better reconstruction overall. Even though this selection method is quite coarse, in our experience, it is a lot more stable than tighter selection methods, while delivering equivalent results. Furthermore, we do want to keep enough geometry in the vicinity of the defect's border, such that the meshes can translate, rotate and slide during the alignment, while still having enough overlap such that good correspondences can be found.
In the second step, we convert the aligned skull templates into voxel space, where we then combine them into a single voxel grid G. This grid G then indicates the ratio of the total number of template skulls that are active at a given voxel position, whereas 'active' can be understood as 'voxels that represent solid bone material, as opposed to empty space. By then applying a threshold T to the grid G, we can extract regions (voxels with G $>$ T), where there is a strong agreement between the templates that this area contains bone material.
However, at this stage, the implant usually contains a lot of unwanted geometry beyond the area of the defect. To get rid of it, one can subtract the damaged target from the reconstruction, which then removes most of this false positive geometry. Depending on the chosen threshold and the quality of the alignment, there might exist areas where the thresholded geometry extends beyond the outer surface of the damaged skull, which is therefore not removed after subtraction, as seen in the top right of Figure \ref{fig:grid_operators}.
To remove those artifacts, the user can offset the target skull along the surface normals before executing the subtraction. By providing a user selection of the defect's border inside, we can additionally set the offset in this area to 0 to avoid a gap between the target and the final implant. As a last step, various filters, which we will refer to as grid operators, are applied for a final cleanup. Each grid operator accepts a list of input grids to be processed and returns a list of output grids, which then serve as an input for the subsequent operations. Although each grid operator in itself is usually quite simple, they can, when combined correctly, improve the result significantly, as demonstrated in Figure \ref{fig:grid_operators}. Currently, there are three operators available. The 'Segment SDF' operator splits the volume into disconnected geometry islands. The 'Filter Grids' operator then allows to select only a certain subset, such as the first \textit{n} grids, while the 'Level Set Filter' operator allows to apply various kinds of smoothing operations. At the end of this process, the voxel space representation is then converted back to a mesh and the final implant can then be exported.

\section{Experiments and Results}
\label{sec:results}

\subsection{Data}
For the following results, we took advantage of the training dataset that was created as part of the AutoImplant challenge \cite{autoimplant}. It is comprised of more than 100 complete skulls extracted from the CQ500 dataset, and about 500 derived variations with artificially induced artifacts in conjunction with their corresponding ground truth reference implant \cite{autoimplant}.

\subsection{Experiment}
To create the implants, we first load the target skull into the scene, adjust the position of the region of interest selection, and then offset the target skull along its normals. This setup then serves as the basis for the four possible test combinations between the tow clipping options and two template selection methods, though slight adjustments to the position of the region picker or target offset have been made where needed. All implants were created under 15 minutes.
The test case number used in Table \ref{tab:results_deviation} and Figures \ref{fig:clipping_comparison}-\ref{fig:signed_distance} is comprised as follows. The first letter indicates the subdirectory the target was taken from (\textbf{f}rontoorbital or \textbf{b}ilateral) followed by the file number. The 'A' indicates that \textbf{a}ll healthy skulls have been added to the scene, aligned to the target, and only the 20 best ones (according to the ICP alignment score) are kept, while 'S' indicates that 20 \textbf{s}imilar skulls have been selected from the dataset with the GASD feature descriptor. The last letter, 'N' or 'C' stands for \textbf{n}o clipping and \textbf{c}lipping.

\subsection{Results}
\label{sec:implant_evaluation}

\begin{table}[htb]
    \centering
    \begin{tabular}{ |c|c|c|c|c|c| } 
     \hline
        Testcase & Min & Max & MAE & HD95 \\ \hline
        B000AN &  -1.3316 & 2.65437 & 0.713973 &  2.36949\\
        B000AC &  -2.11239 & 2.39526 & 0.599912 & 1.90234\\
        B000SN &  -34.3889 &  3.07442 & 5.07532 & 24.88\\
        B000SC &  -2.40375 &  2.31908 & 0.457325 & 1.68154\\
        F010AN & -1.64093 & 2.34638 & 0.518697 & 4.28153\\
        F010AC & -11.1606 & 1.74465 & 1.75776 & 8.50098\\
        F010SN & -3.88612 & 2.16038 & 0.614255 & 3.03085\\
        F010SC & -6.28083 & 1.89999 & 1.14099 & 4.68187\\
     \hline
    \end{tabular}
    \caption{Signed distance from target, measured in mm.}
    \label{tab:results_deviation}
\end{table}

\begin{figure}[htb]
\begin{minipage}[b]{.48\linewidth}
  \centering
  \centerline{\includegraphics[width=3.8cm]{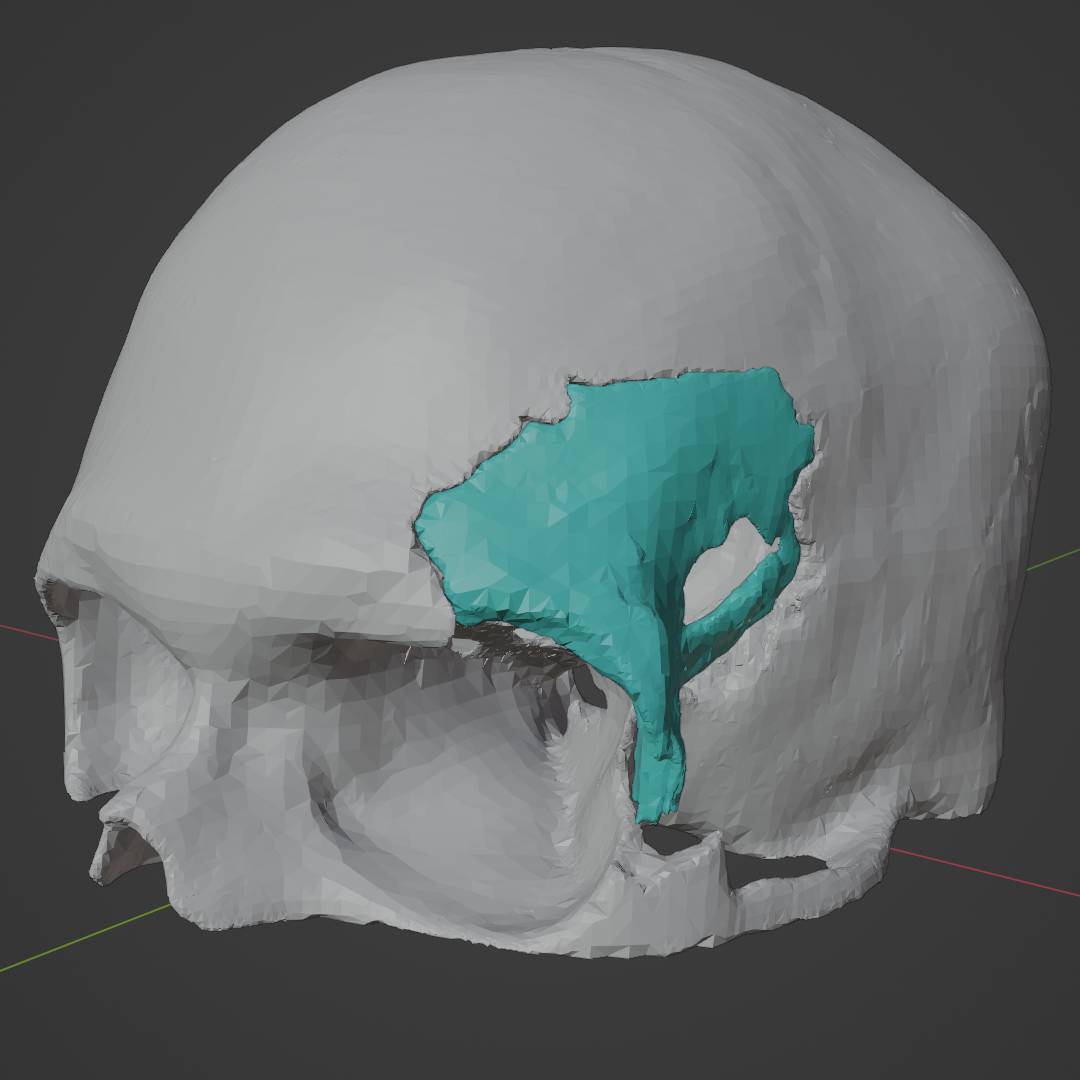}}
  \centerline{(a) F010SN (no clipping)}\medskip
\end{minipage}
\hfill
\begin{minipage}[b]{0.48\linewidth}
  \centering
  \centerline{\includegraphics[width=3.8cm]{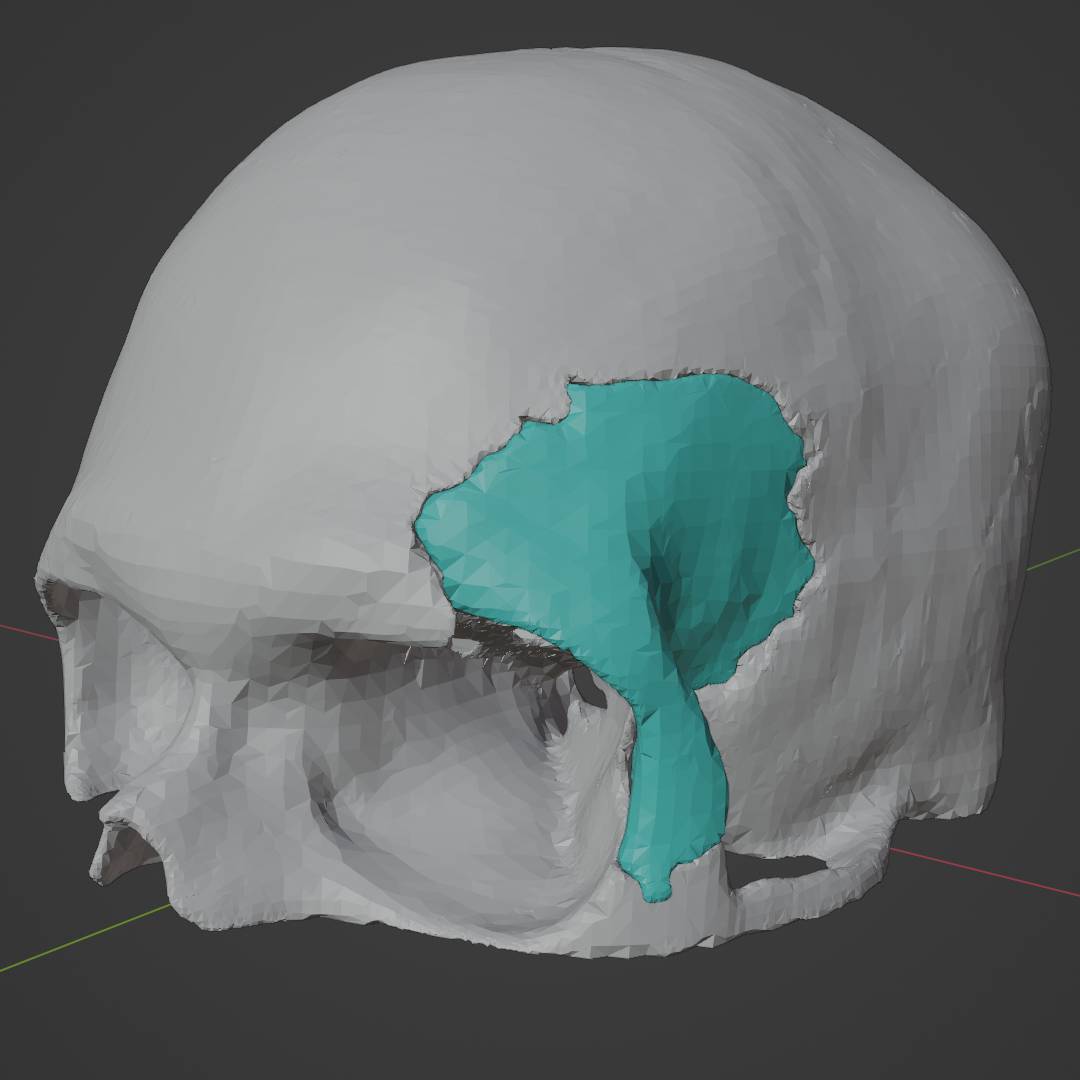}}
  \centerline{(b) F010SC (with clipping)}\medskip
\end{minipage}
\caption{Comparison of reconstructions options with (b) and without (a) clipping.}
\label{fig:clipping_comparison}
\end{figure}

\begin{figure}[htb]
\begin{minipage}[b]{.48\linewidth}
  \centering
  \centerline{\includegraphics[width=3.8cm]{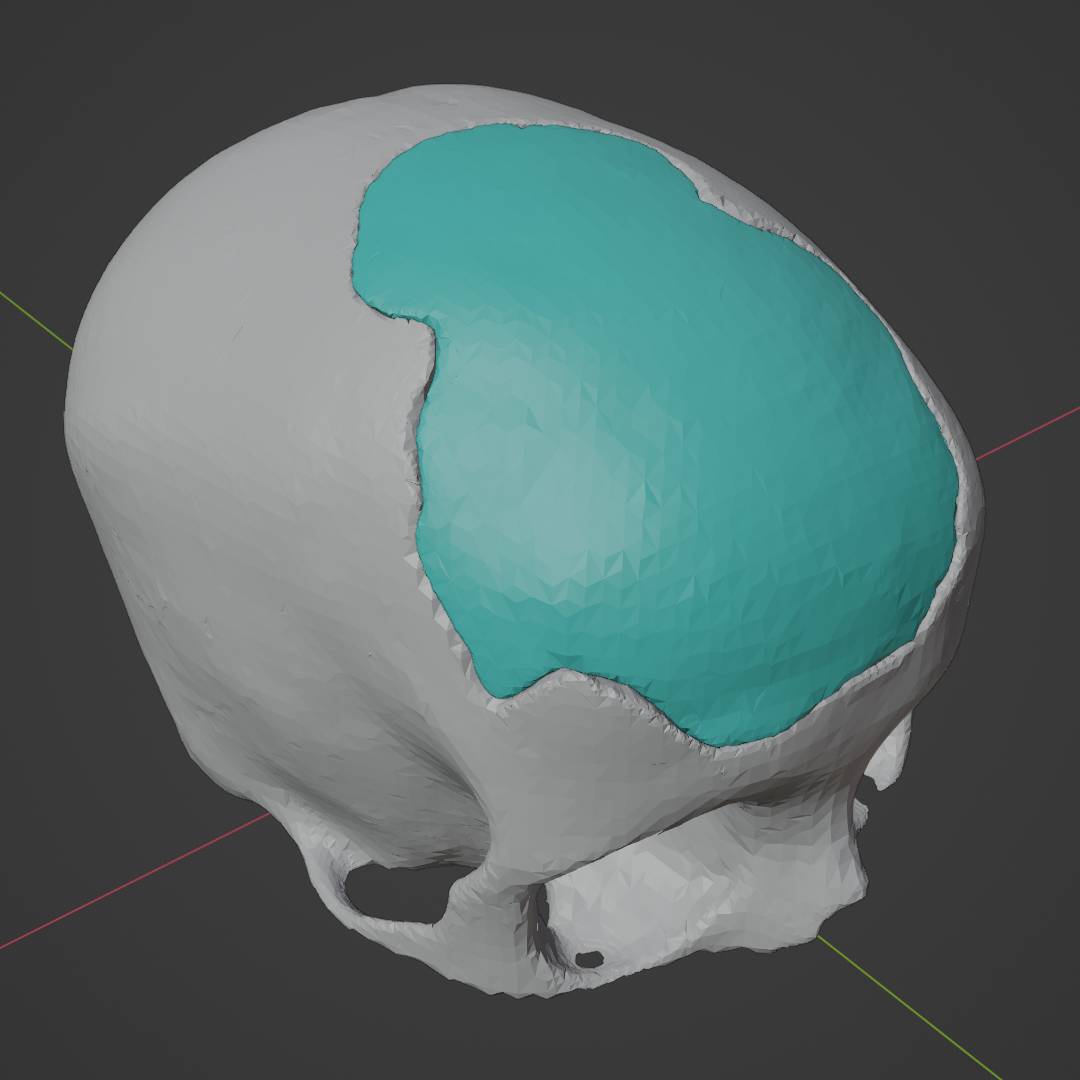}}
  \centerline{(a) B000SC outside}\medskip
\end{minipage}
\hfill
\begin{minipage}[b]{0.48\linewidth}
  \centering
  \centerline{\includegraphics[width=3.8cm]{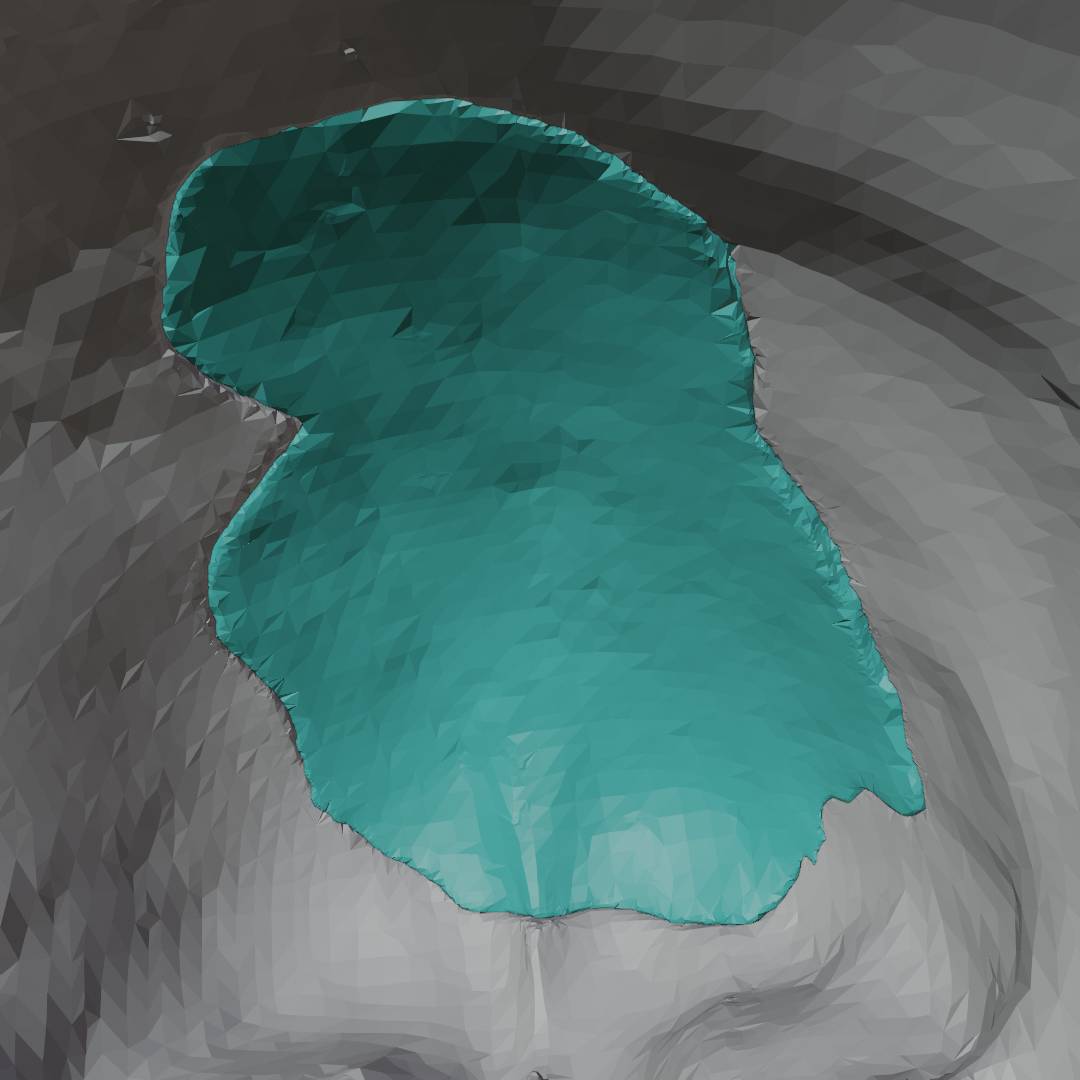}}
  \centerline{(b) B000SC inside}\medskip
\end{minipage}
\caption{A reconstruction of the bilateral region viewed from outside (a) and inside (b) of the skull. }
\label{fig:reconstruction_bilateral}
\end{figure}

\begin{figure}[htb]
\begin{minipage}[b]{.48\linewidth}
  \centering
  \centerline{\includegraphics[width=3.8cm]{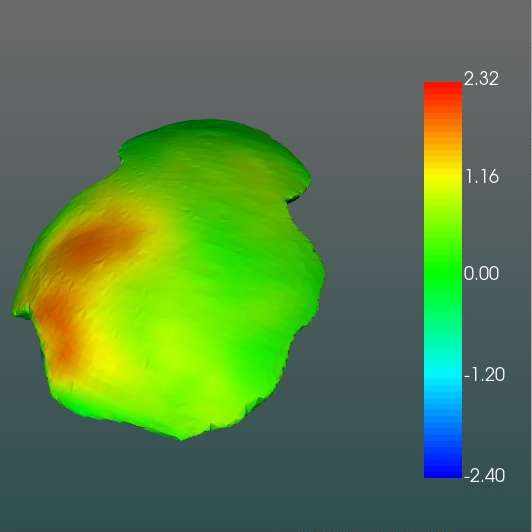}}
  \centerline{(a) B000SC outside}\medskip
\end{minipage}
\hfill
\begin{minipage}[b]{0.48\linewidth}
  \centering
  \centerline{\includegraphics[width=3.8cm]{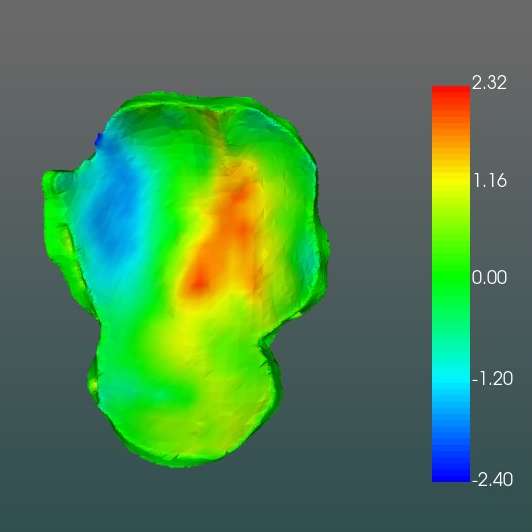}}
  \centerline{(b) B000SC inside}\medskip
\end{minipage}
\caption{Signed distance colormap (a,b) between the reconstructed implant and the ground truth. Green indicates a distance of 0. }
\label{fig:signed_distance}
\end{figure}

\begin{figure}[htb]
\begin{minipage}[b]{.48\linewidth}
  \centering
  \centerline{\includegraphics[width=3.8cm]{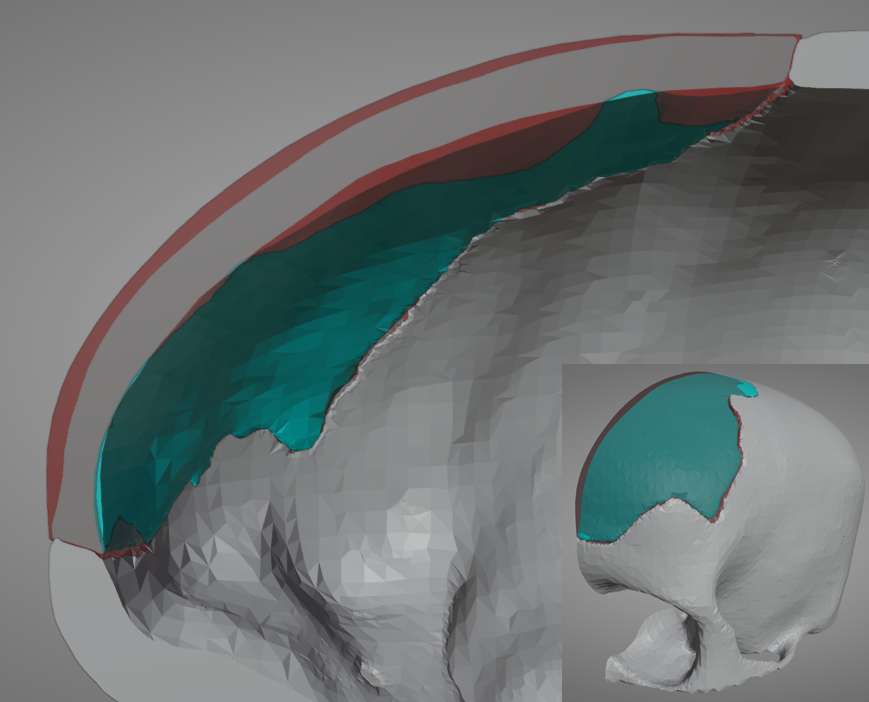}}
  \centerline{(a) Acceptable deviations.}\medskip
\end{minipage}
\hfill
\begin{minipage}[b]{0.48\linewidth}
  \centering
  \centerline{\includegraphics[width=3.8cm]{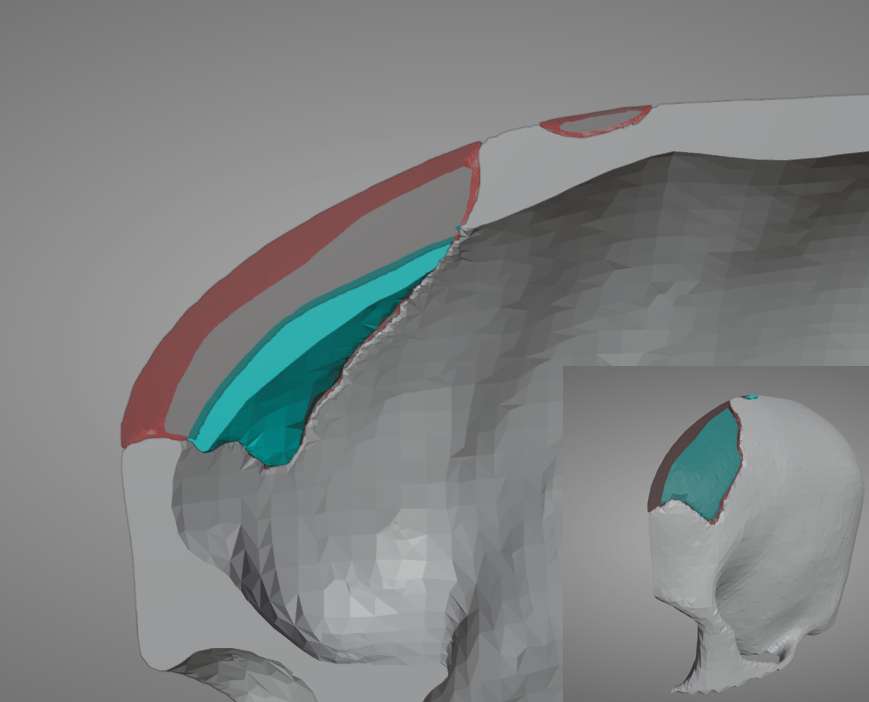}}
  \centerline{(b) Severe deviations.}\medskip
\end{minipage}
\caption{A cross section, showing the deviations between the ground truth (red) and the reconstructed implant (turquoise). (a) and (b) show an acceptable and a severe deviation, respectively.}
\label{fig:TB000SC_crosssection}
\end{figure}

Based on the distance scores in Table \ref{tab:results_deviation}, all variations seem to produce similar results, apart from B000SN, where false positive geometry in the frontoorbital region results in large errors. However, the visual comparison, as seen in Figure \ref{fig:clipping_comparison}, reveals noticeable differences in terms of quality. The implants generated with no clipping are worse than when clipping is enabled. In general, we observe the surface of the implants without clipping is in many cases a lot more jagged and the misalignment at the borders of the defect is, in general, larger. Overall, the results with clipping enabled look quite promising. The general shape of the area to reconstruct was captured quite well and the deviation on the inside, as seen in the Figures \ref{fig:reconstruction_bilateral} (b) and Figure \ref{fig:signed_distance} (b), is quite small for most variations. However, the transition between the implant's and existing skull's surface is not optimal, as the implant tends to be below the target's surface and therefore causes a small step at the border, as seen in Figure \ref{fig:reconstruction_bilateral} (a). In general, we observed that the ICP algorithm has a tendency to produce non-tangential transitions, which is to be expected, as the error objective of the algorithm does not include an explicit concept of tangentiality. Furthermore, the quality of the alignment often decreases, as the complexity of the defect's shape and size increases. While the alignment result is usually quite good in areas with homogeneous curvature directions, as found in the parietal and upper frontal regions, it fails to consider small, local deviations in position and curvature. In these cases, the template often aligns well in one particular area of the defect while being misaligned at another, which is most likely due to the fact that the chosen ICP variant is rigid.

Furthermore, the ideal implant thickness might be less than 50\% of the original bone's thickness \cite{li2023towards}. Therefore, small deviations, as seen in Figure \ref{fig:TB000SC_crosssection} (a), are within acceptable limits, as long as the implant's surface does not penetrate the brain's volume. However, due to the limitations of the rigid ICP algorithm mentioned previously, we usually also have penetrating local areas, as seen in Figure \ref{fig:TB000SC_crosssection} (b). While these kinds of intersections with the brain could be partially avoided by choosing a slightly higher thresholding value T, which in turn results in a thinner implant, the lack of cross-sectional views and other visualization methods within our application makes it usually quite hard to identify these problematic regions during reconstruction. 

\section{Discussion}
The quality of the implants generated using our method is stable regardless of the size, shape, and location of the defects, while for fully data-driven approaches, the results depend heavily on these factors \cite{AutomaticSkullDefect, autoimplant, li2023towards}. Besides, many of the data-driven methods are not optimized in terms of computational efficiency and generalizability \cite{autoimplant,autoimplant2,AutomaticSkullDefect}.
Therefore, it is worth pointing out that the optional GASD feature descriptor is the only part of our application that uses the global shape information of the skulls. This leads to the interesting observation that we do not need the entire skull as part of the reconstruction, but rather only a local, bounded neighborhood in the vicinity of the defect. While this insight is not particularly useful for our CPU-bound application, where plenty of memory is available, it is highly interesting for GPU-bound, machine learning models, as they could potentially save a lot of memory and training time by excluding the geometry outside of the defect's local neighborhood, as partly demonstrated by the work of Li et al. \cite{autoimplant_local_shape}.

Another aspect to consider is the fact that fully data-driven approaches trade automatization for a lack of manual intervention, which might be needed in cases where the method is not able to produce the desired result. For example, Ellis et al. \cite{qualitativeCriteria} states that almost all of the evaluated implants exceeded the ideal thickness of 50\%, which in turn made them unsuitable for cranioplasty \cite{qualitativeCriteria}. Furthermore, many data-driven approaches do not generalize well to real-world defects out of the box, as they are usually trained on skulls with artificially induced defects \cite{li2023towards}. However, in terms of convenience, the data-driven methods outperform semi-automatic approaches, as demonstrated by the website called 'StudierFenster' \cite{studierfenster}, where the user only needs to select the input skull to be reconstructed.

Despite some of the challenges mentioned, data-driven methods, as proposed in \cite{AutomaticSkullDefect, autoimplant, li2023towards}, do produce state-of-the-art results nevertheless. In the evaluation performed as part of the AutoImplant 2021 Challenge \cite{qualitativeCriteria}, experts assessed four submissions based on \textit{false positive area}, \textit{completeness}, \textit{fit}, and \textit{overall feasibility} \cite{qualitativeCriteria}. Similarly to our results, all submissions of the study contain 'minimal' to 'gross' amounts of false positive geometry. In terms of\textit{ completeness}, the majority of implants perform well \cite{qualitativeCriteria}. In our application, we observed that the \textit{completeness} depends on the quality of the alignment, location of the defect, and threshold value used. However, in most cases tested so far, we were able to fix these by simply adjusting some settings. Similar to the data-driven approaches, our method struggles to achieve a good \textit{fit}, especially at the borders, which describes the overall match in shape and profile between the target and the implant \cite{li2023towards, qualitativeCriteria}.

\section{Conclusion}
In this paper, we presented a semi-automatic solution to the cranial implant design problem, by developing a software prototype capable of generating reasonable implants with only a few high-level user interactions. Furthermore, we demonstrated that clipping of geometry can improve the results significantly. The degree of automatization of the proposed semi-automatic method still needs to be further increased, and issues, like the inconsistencies at the implant's borders and the local penetrating areas, need to be addressed in the future by using more advanced alignment and post-processing methods.


\section{Acknowledgments}
\label{sec:acknowledgments}
The work is supported by the REACT-EU project KITE (EFRE-0801977), ”NUM 2.0“ (FKZ: 01KX2121) and FWF enFaced 2.0 (KLI 1044). 

\bibliographystyle{IEEEbib}
\bibliography{strings,refs}

\end{document}